\documentclass[fleqn,usenatbib]{mnras}

\usepackage{newtxtext,newtxmath}

\usepackage[T1]{fontenc}

\DeclareRobustCommand{\VAN}[3]{#2}
\let\VANthebibliography\thebibliography
\def\thebibliography{\DeclareRobustCommand{\VAN}[3]{##3}\VANthebibliography}

\usepackage{graphicx}	
\usepackage{amsmath}	

\title[The double quasar Q2138-431]
 {The double quasar Q2138-431: detection of a lensing galaxy}

\author[M. R. S. Hawkins]{
M. R. S. Hawkins $^{1}$\thanks{E-mail: mrsh@roe.ac.uk}
\\
$^{1}$Institute for Astronomy (IfA), University of Edinburgh,
 Royal Observatory, Blackford Hill, Edinburgh EH9 3HJ, UK\\}

\date{Accepted XXX. Received YYY; in original form ZZZ}

\pubyear{2021}

\begin{document}
\label{firstpage}
\pagerange{\pageref{firstpage}--\pageref{lastpage}}
\maketitle

\begin{abstract}
This paper reviews the question of whether the wide separation double
quasar Q2138-431 is a gravitational lens.  From early work, the two quasar
images are known to have almost identical spectra and redshifts, but no
lensing galaxy has so far been detected.  In this paper we used recent
deep surveys in infrared and optical bands to search for the presence of a
galaxy with the expected properties of a gravitational lens. The search
revealed a $5 \sigma$ detection of a faint galaxy between the two quasar
images on a deep $J$-band frame from the VISTA Science Archive, with
apparent magnitude $J = 20.68$.  Non-detection in the $I$-band implied a
redshift $z > 0.6$, and mass modelling of the quasar system gave a mass of
$1.31 \times 10^{12} M_\odot$ for the lensing galaxy, with mass-to-light
ratio $M_{\odot}/L_{\odot} = 9.0$.  Archival photographic data from the UK
1.2m Schmidt telescope covering 25 years were used to construct light
curves for the two quasar images, which were then cross-correlated to
measure any time lag.  This showed image B to lead image A by around a
year, consistent with 410 days from the mass model. Although the
similarity of the spectra and the detection of the lensing galaxy are the
most compelling arguments for the classification of Q2138-431 as a
gravitational lens, the time delay and mass-to-light ratio provide a
consistent picture to support this conclusion.  The wide separation of the
quasar images and the simplicity of the mass model make Q2138-431 an
excellent system for the measurement of the Hubble constant.
\end{abstract}




\section{Introduction}

\begin{figure}
\begin{picture}(200,230)(-5,5)
\includegraphics[width=0.46\textwidth]{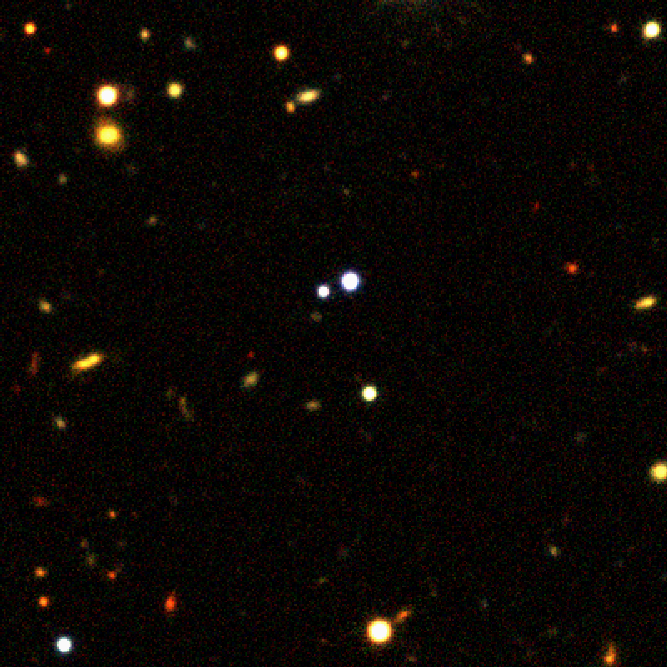}
\end{picture}
\caption{Composite 3-colour image from DES frames in $g$, $r$ and $i$
 passbands, showing the galaxy environment of the two quasar images near
 the centre of the frame, with the brighter image A to the right and
 image B to the left.  The field also includes some of the local
 standard stars used for the measurement of the light curve.  The frame
 is approximately 20 arcsec on a side, and north is up the page, east to
 the left.}
\label{fig1}
\end{figure}

The double quasar Q2138-431 was discovered as part of a photographic
survey for quasars based on optical variability \citep{ha97}.  It was
detected as an elongated blue image which varied by over a magnitude in
a few years, and closer inspection showed it to comprise two star-like
images separated by about 4.5 arcsec.  Subsequent analysis, which we
briefly review in Section~\ref{obs}, revealed that the two images were
quasars with the same redshift, $z = 1.641$.  This raised the possibility
that the quasar images were part of a gravitational lens system, although
there was no obvious sign of a lensing galaxy.  In Fig.~\ref{fig1} we have
constructed a composite 3-colour image using frames from the Dark Energy
Survey (DES) \citep{a18} in the $g$, $r$ and $i$ passbands.  The two
quasar images are clearly visible near the centre, together with a sparse
population of mostly red galaxies with a limiting magnitude of
$R \approx 24$.

Wide separation gravitational lenses with image separations greater than
two arsecs are rare, and have proved particularly useful in investigating
dark matter distributions \citep{f10}, and the measurement of the Hubble
constant \citep{e05,s13,w17}.  Should Q2138-431 be shown to be a
two-component gravitational lens, it would be among the widest separation
systems known, and very well suited to mass modelling and the measurement
of a time delay for calculating the Hubble constant.  In the
CASTLES\footnote{https://www.cfa.harvard.edu/castles/} database of
gravitational lenses, there are only five systems with separations greater
than 4 arcsec.  Two of these are quadruple cluster lenses
\citep{o08,f08,h20}, which although interesting in their own right
present major difficulties for the measurement of the Hubble constant.
The remaining three are doubles, of which RXJ 0921+4529 has now been shown
to be a binary system \citep{p10}.  CLASS B2108+213, which was discovered
as a double lensed radio source with a smooth optical spectrum, is
tentatively identified as a BL Lac object \citep{m05,m10}.  So far the
source redshift has not been measured, and the complexity of the lensing
group or cluster suggests some difficulty in measuring the value of the
Hubble constant.  The final large separation lens is the well-known double
Q0957+561 \citep{w79} which has indeed proved useful for the measurement
of the Hubble constant \citep{f10}, but even here the asymmetry of the
positions of the quasar images and galaxy lenses has complicated the
modelling of the system.

In the paper reporting the discovery of the double quasar Q2138-431,
\cite{ha97} showed that the spectra of the two quasar images are almost
identical, and that the redshifts are the same to within very small
errors.  This made a strong case for the two images to be gravitationally
lensed, but long CCD integrations in the $R$-band failed to reveal a
lensing galaxy. For this paper we have taken advantage of modern deep
photometric surveys, especially in the infrared, to look again for a
lensing galaxy between the two quasar images.  Image subtraction of the
quasar images has resulted in the positive $5 \sigma$ detection of a
galaxy in the $J$-band, but no significant detections in the $r$-, $i$-
or $z$-bands.  We used the photometry in these passbands to estimate a
lower limit to the redshift of the lensing galaxy, and hence its absolute
magnitude.  We also used astrometric measurements of the positions of the
quasar images and lensing galaxy to fit a mass model to the system, and
obtain a value for the mass of the deflector.  The resulting parameters
imply that the lens is a massive luminous red galaxy (LRG).  The survey
data on which this analysis is based is described in Section~\ref{obs},
and the image analysis techniques and photometry in Section~\ref{gal}.

For a gravitational lens system to be useful for the measurement of the
Hubble constant, it is necessary to measure time delays between the light
paths to the images.  The photographic survey which resulted in the
discovery of Q2138-431 is described in more detail in Section~\ref{obs},
and comprised long runs of yearly observations in several passbands
between the years 1977 and 2001.  These plates yield light curves for the
two quasar images which can be cross-correlated to provide a preliminary
estimate of the time lag between variations in the two images, which we
describe in Section~\ref{lc}.  We then compare this result with the
predicted time delay from the mass model.  Throughout the paper we assume
$\Omega_M = 0.27$, $\Omega_\Lambda = 0.73$ and
$H_0 = 71$ km s$^{-1}$ Mpc$^{-1}$ for cosmological calculations.

\section{Observations}
\label{obs}

The double quasar Q2138-431 was discovered as part of a survey for
variable objects with an ultra-violet excess and an elongated image on
photographic plates from the 1.2m UK Schmidt telescope at Siding Springs
Observatory, Australia \citep{ha97}.  The purpose of the survey was to
find gravitationally lensed quasars, and Q2138-431 was the most promising
candidate which on closer examination was seen to comprise two images
separated by 4.5 arcsec.  Spectroscopy of the two images was obtained with
the ESO 3.6m telescope at La Silla, Chile, which showed them to be quasars 
with no detectable difference between the two high signal-to-noise
spectra, as illustrated in Figure 2 of \cite{ha97}.  The redshifts of the
two quasar images were the same at $z = 1.641$, and cross-correlation
between the two spectra showed a difference in velocity of
$0 \pm 114$ km s$^{-1}$.  Such a small velocity difference is hard to
account for in a random alignment or binary pair, and strongly suggested
that the quasar images were part of a gravitational lens system.  There
was however no obvious evidence for a lensing galaxy.  In order to clarify
the situation, deep CCD observations of the quasar images were obtained in
the $R$-band, and the area between the two quasar images investigated
using image subtraction techniques to reduce contamination from the quasar
light.  No lensing galaxy was detected, with a lower limit of $R > 23.8$.

The failure to find a lensing galaxy suggested the possibility that the
lens might be a dark galaxy, an idea which was receiving some attention at
that time \citep{h97}.  However, the limit obtained in the $R$-band is
only useful for relatively low redshift lenses.  The 4000 \AA\ feature
starts to leave the $R$-band at around $z \approx 0.4$ and from the
$I$-band at $z \approx 0.8$, making the detection of a lensing galaxy in
this regime very challenging.  On this basis we moved the search into
the infrared, and made use of the $J$-band survey from the VISTA Science
Archive\footnote{http://horus.roe.ac.uk/vsa/} (VSA).  The archive contains
image tiles which we used for the detection of the lensing galaxy, and
catalogue data which we used for calibration.  The details of the
analysis are described in Section~\ref{gal}.  In addition to the
$J$-band data, we also made use of $r$-, $i$- and $z$-band data from the
Dark Energy Survey\footnote{https://des.ncsa.illinois.edu/releases/dr1}
(DES) \citep{a18} to look for detections or upper limits in the reddest
available optical bands.  Data from this survey were also used for the
construction of a local sequence to calibrate the light curves described
in Section~\ref{lc}.  The light curves themselves were measured from an
extension of the series of photographic plates used in the discovery of
Q2138-431, and the SuperCOSMOS scans form part of the SuperCOSMOS Science
Archive\footnote{http://ssa.roe.ac.uk} (SSA).  The measurement and
analysis of the light curves of the two quasar images is described in
detail in Section~\ref{lc}.

\section{The Lensing galaxy}
\label{gal}

\subsection{Infrared observations}

\begin{figure*}
\centering
\begin{picture} (0,280) (255,0)
\includegraphics[width=0.49\textwidth]{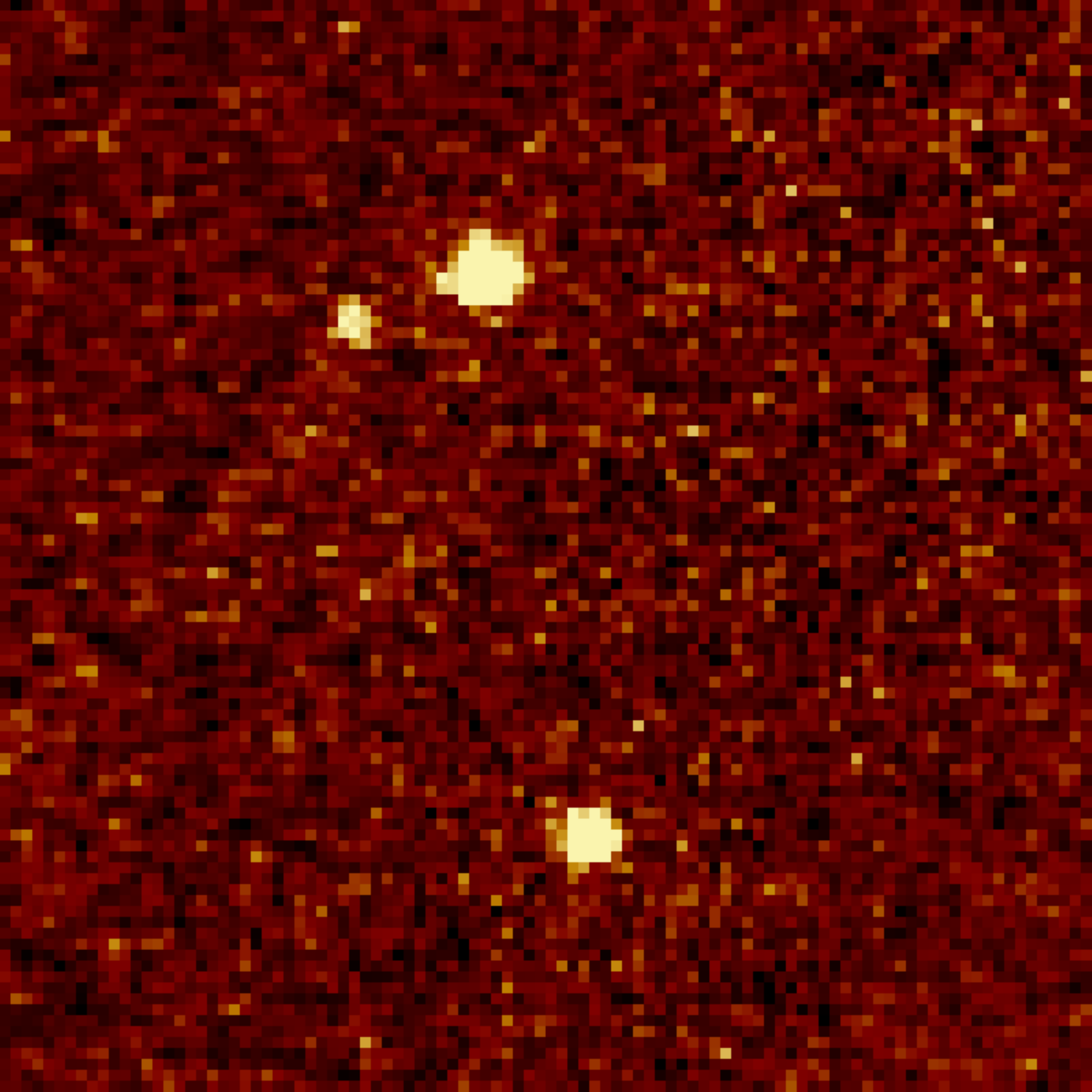}
\end{picture}
\begin{picture} (0,280) (-5,0)
\includegraphics[width=0.49\textwidth]{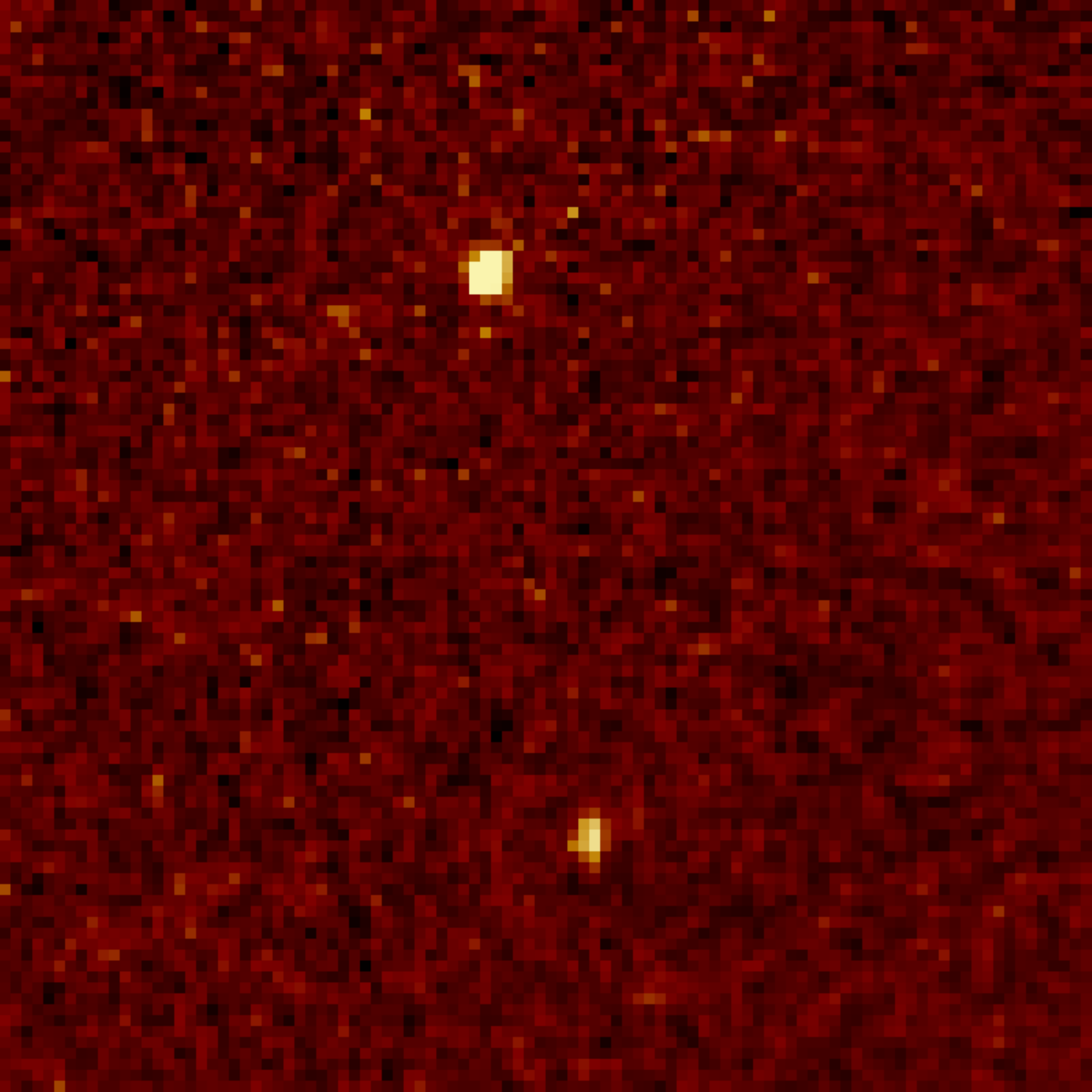}
\end{picture}
\caption{Frames from the VISTA Science Archive in the $J$-band
(left hand panel) and $K_s$-band (right hand panel).  The frames are
 approximately 30 arcsec on a side, and north is up the page, east to the
 left.  The frames show the double image of Q2138-431 at the top, with
 image A to the right and image B to the left, and a nearby star used to
 fit the PSF below.}
\label{fig2}
\end{figure*}

The Q2138-431 system has celestial coordinates $21^{\rm h}$ $41^{\rm m}$
$16.27^{\rm s}$, -42$^{\circ}$ $57^{\prime}$ $10.0^{\prime\prime}$ (2000)
and in the VISTA Science Archive is covered by frames in the $J$
and $K_s$ passbands.  The frames have a pixel size of 0.341 arcsec,
and median seeing FWHM = 1.18 arcsec.  In Fig.~\ref{fig2} we show cutouts
from these frames, with the double quasar at the top, and below it a star
which has been helpful in the photometric calibration and for image
subtraction.  Cursory examination of the $J$-band image in the left hand
panel shows a suggestive bulge along the line of centres of the two quasar
images, which has the potential to be the lensing galaxy.  The $K_s$-band
image in the right hand panel of Fig.~\ref{fig2} does not show this
feature, but the frame does not go very deep, and indeed image B of the
double quasar is barely visible.  We discuss detection limits in detail
below.

In order to investigate the possibilty that the feature between the quasar
images is the lensing galaxy, we undertook a PSF subtraction procedure to
remove the contaminating quasar light.  Attempts to use stars in the field
as models for the PSF, including the one visible in Fig.~\ref{fig2}, were
not very successful due to difficulties in registration and poor
signal-to-noise.  The best approach proved to be to use a Moffat profile
\citep{t01} of the form:

\begin{equation}
P(r) = h \biggl[ 1+\Bigl(\frac{r}{\alpha}\Bigr)^2 \biggr]^{-\beta}
\label{eqn1}
\end{equation}

\noindent fitted to a star of similar magnitude to the quasar images.  We
first rebinned the array to improve registration giving the images shown
in the left hand panel of Fig.~\ref{fig3}, and then measured the image
centroids with the Starlink GAIA package.  The next step was to fit the
Moffat profile to the star in Fig.~\ref{fig2}.  We followed \cite{t01} and
set $\beta = 4.765$, and then varied $\alpha$ and $h$ to give the best
fit.  For the PSF subtraction we kept these values of $\alpha$ and $\beta$
and for each of the two quasar images varied $h$ to get the best fit.  The
resulting profiles $P(r)$ were then subtracted from each image using the
measured centroids.  The results of this procedure are shown in the right
hand panel of Fig.~\ref{fig3}.  Contours of the subtracted quasar images
are included to show the location and shape of the galaxy image.

In order to assess the likelihood that we had a detection of the lensing
galaxy, the first step was to measure its brightness.  To do this, we
used local photometric standards from the VSA catalogue, and then with
photometry routines from the GAIA package we measured the apparent
magnitude of the lensing galaxy to be $J = 20.68 \pm 0.09$.  Based on the
rms variation of the sky background, this is a $5 \sigma$ detection.  The
detection limit of the frame was determined from the faintest $5 \sigma$
detections listed in the VSA catalogue, and verified by our photometry of
the faintest images observed on the frame covering the quasar position.
The $5 \sigma$ detection limit we derived was $J \lesssim 20.7$.

The right hand panel of Fig.~\ref{fig2} shows the quasar system on the
$K_s$ frame from the VSA survey.  Visual inspection of the area between
the two quasar images shows no evidence for the presence of a lensing
galaxy.  The $K_s$ frame is clearly less deep than the one for the
$J$-band, and the B image of the quasar is very faint, but we found it
useful to measure the detection limit to put bounds on the colour of the
lensing galaxy and its redshift.  Using the same approach as for the
$J$-band we found the detection limit to be $K_s \lesssim 18.7$.  To
estimate the $K_s$ magnitude of the lensing galaxy we used the intrinsic
colours for luminous red galaxies (LRGs) from \cite{m01a}.  For the $J-K$
passband they give $J-K = 0.87$, which implies $K_s \gtrsim 19.8$, well
below the detection limit of the $K_s$ frame and consistent with
non-detection in the $K_s$ passband.  Applying $K$ and evolutionary
corrections from \cite{p97} up to a redshift $z \sim 1$ does not
significantly alter this.

\subsection{Optical observations}

In addition to the $J$- and $K_s$-band images from the VSA survey, the
Q2138-431 system was also observed as part of the Dark Energy Survey
in the $g$, $r$, $i$ and $z$ passbands \citep{a18}.  These frames
have a pixel size of 0.263 arcsec with a median seeing FWHM = 0.94
arcsec, and proved very useful in putting limits on the redshift of the
lensing galaxy.  To estimate the expected magnitudes of the lensing
galaxy in these passbands, we again made use of the intrinsic colours of
\cite{m01a}, based on template spectra of a large sample of galaxies.  We
converted the Johnson colours of \cite{m01a} to the SDSS system using the
colour transformations of \cite{j06}.  These were then applied to the
observed $J$ magnitude of the lensing galaxy to obtain estimated
magnitudes of $g = 23.56$, $r = 22.73$, $i = 22.36$ and $z = 22.18$ in the
galaxy rest frame.  Examination of the DES frames showed no indication of
the presence of a lensing galaxy, in agreement with the results of
\cite{ha97} from CCD frames of similar depth to the DES observations.
The 95\% completeness limits in these passbands \citep{a18} are
$g < 23.72$, $r < 23.35$, $i < 22.88$ and $z < 22.25$.  In all cases the
estimated rest wavelength magnitudes are within these bounds, and so the
non-detections can be used to set minimum values for the $K$ corrections,
and hence the minimum redshift of the lensing galaxy.  As might be
expected, the non-detection in the $i$-band provides the strongest
constraint on the redshift of the lensing galaxy.  Using $K$ corrections
from \cite{p97}, the minimum redshift for the lensing galaxy to move below
the detection threshhold is $z \gtrsim 0.5$.

So far, we have neglected the possible effect of an evolutionary
correction in our calculations.  Evolutionary corrections can be
unreliable and model dependent \citep{d15}, but this can be allowed for by
comparing the observed lower limit to the $i-J$ colour with the intrinsic
$(i-J)_0$ colour.  In this case, the large evolutionary corrections
largely cancel out, and the redshift can be estimated by determining the
point at which the addition of the combined $K + e$ correction to the
intrinsic colour matches the observation.  We find that
$(i-J)_{obs} > 2.2$, and $(i-J)_0 = 1.7$ implying a $K+e$ correction of
0.5, corresponding to a redshift of $z > 0.6$.  The inclusion of the
evolutionary correction thus slightly tightens the redshift limit.

A lower limit to the redshift of the lensing galaxy of around
$z \gtrsim 0.6$ is not surprising, given the faintness of the $J$
magnitude, and provides an explanation as to why the lens galaxy is not
detected in optical passbands.  The lower limit on $z$ also implies a
lower limit on the luminosity of the lensing galaxy.  Using the $K$ and
evolutionary corrections for $z = 0.6$ from \cite{p97}, we obtain
$M_J = -21.02$.  Applying a bolometric correction BC = 2.13 for the VISTA
$J$-band frames from \cite{c16} gives an absolute magnitude
$M_{bol} = -23.15$, equivalent to a luminosity
$L_{gal} = 1.45 \times 10^{11} L_\odot$.

\subsection{Lens model}

\begin{figure*}
\centering
\begin{picture} (0,280) (255,0)
\includegraphics[width=0.49\textwidth]{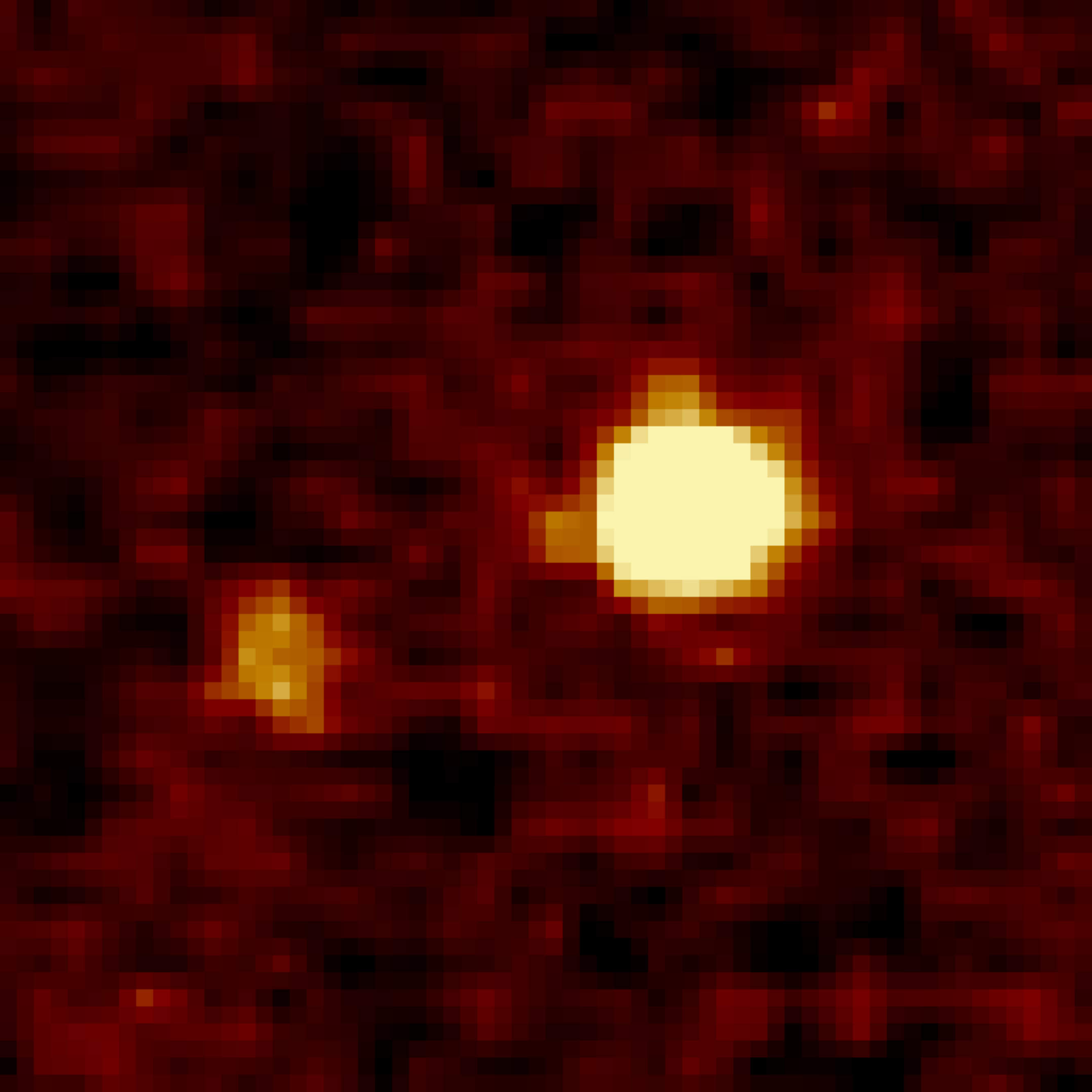}
\end{picture}
\begin{picture} (0,280) (-5,0)
\includegraphics[width=0.49\textwidth]{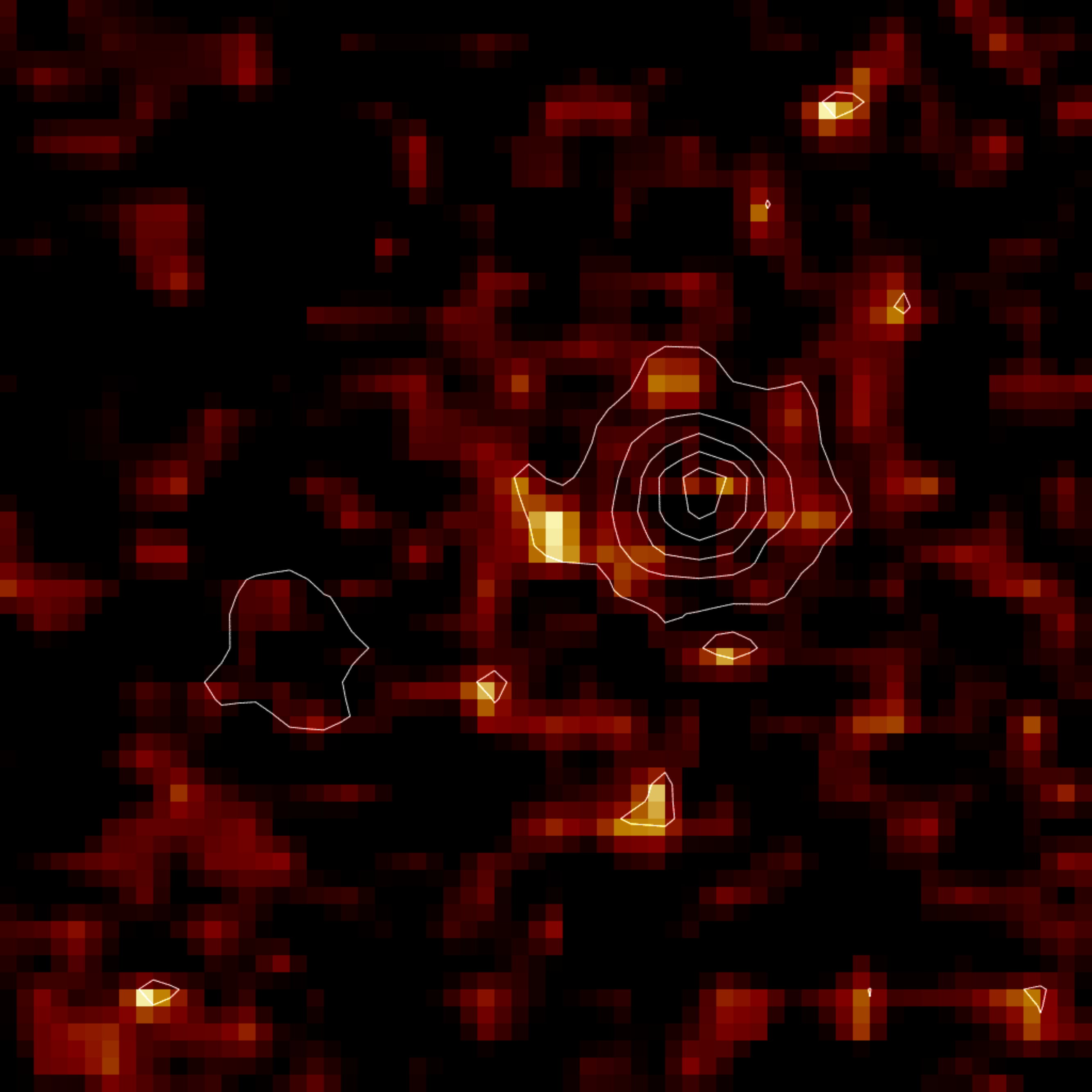}
\end{picture}
\caption{The left hand panel is from a $J$-band frame from the VISTA
 Science Archive.  The frame has been re-binned and shows image A to the
 right and image B to the left.  The small bulge extending to the left of
 image A is the candidate lensing galaxy.  In the right hand panel the
 quasar light has been subtracted, clearly revealing the presence of a
 galaxy between the two quasar images, the positions of which are
 indicated by superimposed contours.  The plots are approximately 10
 arcsec on a side, and north is up the page, east to the left.}
\label{fig3}
\end{figure*}

To provide a consistent picture of the properties of the lensing galaxy,
we made use of the lens modelling software of \cite{k01} to estimate the
lens mass.  We used the GAIA astrometry package to measure the positions
of the two quasar images, obtaining a refined value for the image
separation of 4.481 arcsec, and also the position of the lensing galaxy.
$J$-band flux measures for the two quasar images from the VISTA frame
were also included in the lens model.  We then used routines from the
software of \cite{k01} to fit a singular isothermal ellipsoid plus
external shear $({\rm SIE}+\gamma)$ model to the astrometric and
photometric data, which are summarised in Table~\ref{tab1}.  We obtained
a good fit to the positions, and derived a value for the Einstein radius
of $\theta_E = 2.019$ arcsec, close to that found by \cite {ha97}.  We
obtained $e = 0.1$ for the lens ellipticity, in position angle
-10$^{\circ}$ measured east of north.  For the external shear we found
$\gamma = 0.1$ in position angle 40$^{\circ}$.  The overall goodness of
fit gave $\chi^2 = 12.6$.

\begin{table}
\caption{Summary of the observations for Q2138-431 used in the lens
 modelling.}
\label{tab1}
\begin{center}
\vspace{5mm}
\begin{tabular}{ccccc}
\hline
 \multicolumn{2} {c} {Observations} & A & B & G \\
\hline
 Positions & RA (arsec)  & $4.117 \pm 0.02$ & 0 & $2.815 \pm 0.02$ \\
           & Dec (arsec) & $1.558 \pm 0.02$ & 0 & $1.259 \pm 0.02$ \\
\hline
 Fluxes    & $J$ (mags)          & $18.24 \pm 0.03$ & $19.85 \pm 0.07$ &
 $20.68 \pm 0.09$ \\
\hline
\end{tabular}
\end{center}
\end{table}

The mass $m$ of the lens depends mainly on $\theta_E$, but is also a
function of the quasar and galaxy redshifts.  To be specific,

\begin{equation}
m = \frac{c^2 {\theta_E}^2}{4 G} \frac{D_{LS}}{D_L D_S}
\label{eqn2}
\end{equation}

\noindent where $D_L$, $D_S$ and $D_{LS}$ are the angular diameter
distances to the lens, to the source, and from the lens to the source
respectively.  Using the lower limit $z = 0.6$ derived above, this
gives $m_{lens} = 1.31 \times 10^{12} M_{\odot}$.  Combining this limit
with the luminosity limit $L_{gal} = 1.45 \times 10^{11} L_\odot$ derived
above gives a mass-to-light ratio $M_{\odot}/L_{\odot} = 9.0$.  To give an
idea of the sensitivity of this result to redshift, if we assume $z = 1.0$
then $m_{lens} = 2.88 \times 10^{12} M_{\odot}$,
$L_{gal} = 3.53 \times 10^{11} L_\odot$ and $M_{\odot}/L_{\odot} = 8.2$.

\section{Light curves}
\label{lc}

The discovery of Q2138-431 was part of a long term survey to detect
quasars on the basis of their variability \citep{h96}.  The survey was
based on an extensive sequence of photographic plates in several colours
taken with the UK 1.2m Schmidt Telescope of the ESO/SERC field 287,
centred on $21^{\rm h}$ $28^{\rm m}$, $-45^{\circ}$ (1950).  Of particular
interest for the detection of quasars is the $B_j$ passband, bounded by
the GG395 filter in the blue, and the IIIaJ emulsion cut-off in the red
at about 5400 \AA, and quite close to the SDSS $g$-band.  The plates
were originally measured with the COSMOS measuring machine at the Royal
Observatory, Edinburgh \citep{m84}, which after photometric calibration
produced light curves covering some 16 years.  Although these light curves
were sufficient to detect a large number of quasars from their
variability, including Q2138-431, the machine measures used a low
detection threshhold above the sky, which merged the two quasar images.
This meant that the light curves of the two images could not be measured
separately.  Since the discovery of Q2138-431, the original sequence of
plates in the $B_j$ passband has been extended to include at least one
plate, and usually 4 or more, every year between 1977 and 2001.  These
have now all been measured with the SuperCOSMOS measuring machine
\citep{h01}, and the scans and catalogues form part of the SuperCOSMOS
Science Archive.

\begin{figure}
\centering
\begin{picture} (0,150) (122,00)
\includegraphics[width=0.24\textwidth]{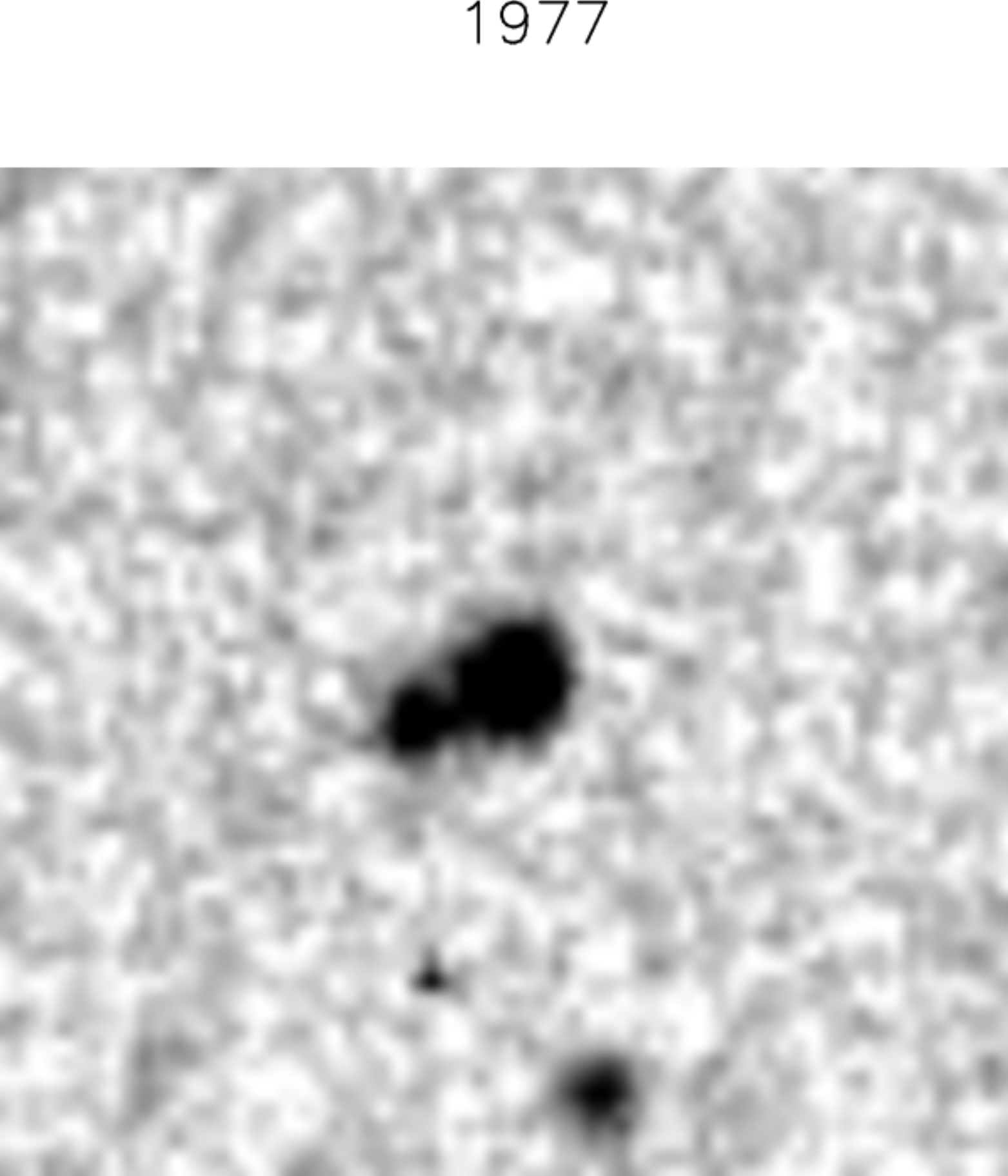}
\end{picture}
\begin{picture} (0,150) (0,00)
\includegraphics[width=0.24\textwidth]{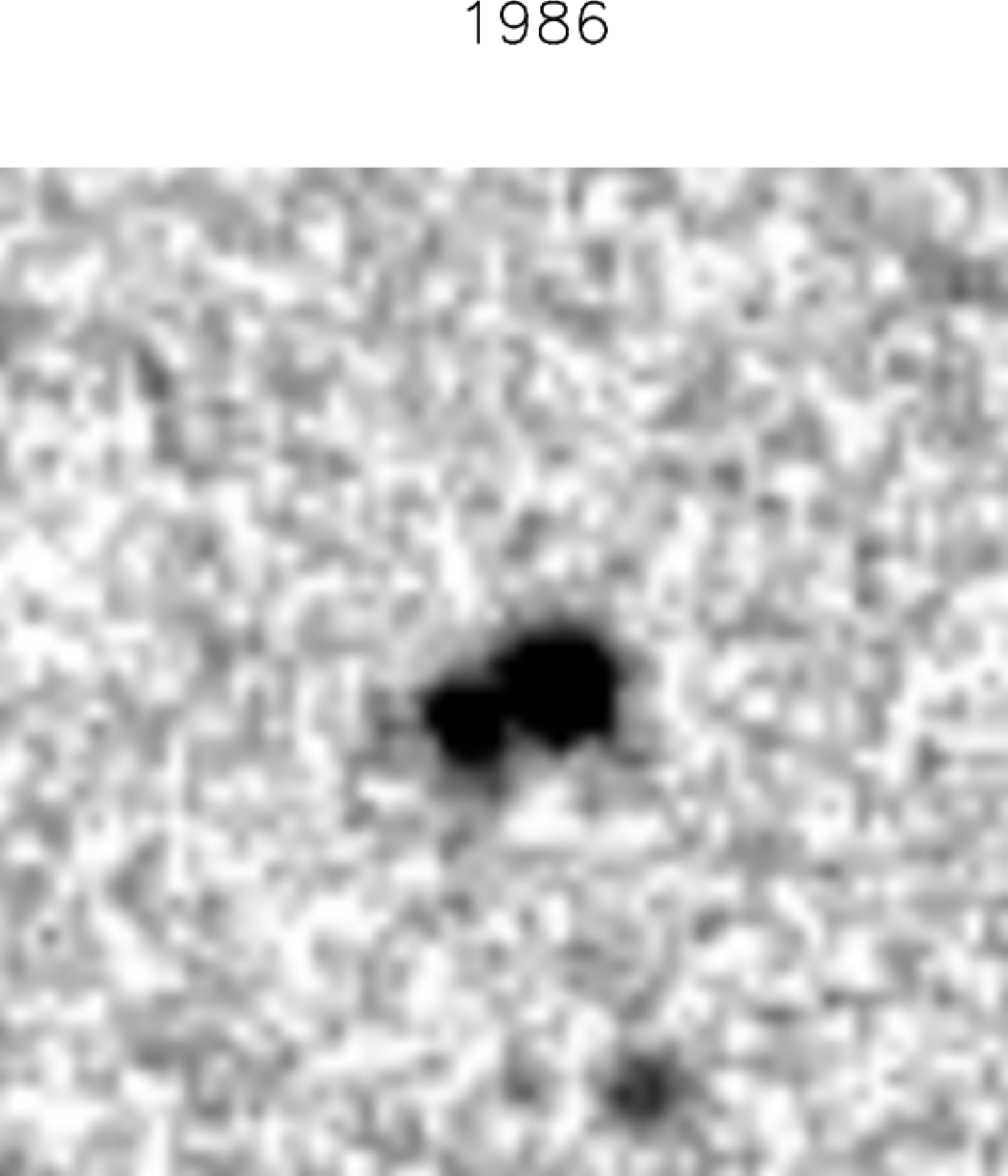}
\end{picture}
\caption{The two frames are from photographic plates taken by the UK 1.2m
 Schmidt telescope, and scanned by the SuperCOSMOS measuring machine at
 the Royal Observatory, Edinburgh.  They illustrate the material on which
 the quasar light curves are based, and the change in magnitude difference
 between the two quasar images in the centre, over a period of 9 years.
 Image A to the right has decreased in brightness by about 0.5 magnitudes
 relative to image B on the left.  Also included near the bottom of the
 plot is a standard star.  The plots are  approximately 40 arcsec on a
 side, and north is up the page, east to the left.}
\label{fig4}
\end{figure}

Although the SuperCOSMOS measures are superior in many ways to the earlier
COSMOS measures, the SSA catalogues are also based on measures with a low
detection threshhold in order to maximize the depth of the survey.  This
again has the effect that the double quasar images are not resolved, and
so we retrieved the original digitized mapping mode scans with a view to
using more sophisticated photometric routines to measure the two images
separately.  Fig.~\ref{fig4} shows cut-outs of the double quasar images
from two of the plates from the quasar variability survey.  The exposures
were taken in 1977 and 1986, and illustrate the change in magnitude
difference between the two images over 9 years.

To measure the light curves we first extracted a square array from the
SuperCOSMOS scan of each plate of side 8 arcmin and centred on
Q2138-431.  We then selected 18 stars in this field, spanning the
magnitudes of the two quasar images, with which to construct a local
photometric sequence.  These stars were then identified in the $g$-band
frame of the area from the DES survey.  Taking advantage of the linearity
of the CCD frames as opposed to the non-linear reciprocity failure of the
photographic plates, we used the GAIA photometric package to measure the
magnitudes of the sequence stars.  The DES $g$-band and the photographic
$B_J$-band are very similar, with an effective wavelength of 4670\AA, and
we could detect no significant colour term.  On this basis we used the
$g$-band measures of the sequence stars to directly calibrate the $B_J$
photographic images.

\begin{figure}
\centering
\begin{picture} (0,590) (120,0)
\includegraphics[width=0.48\textwidth]{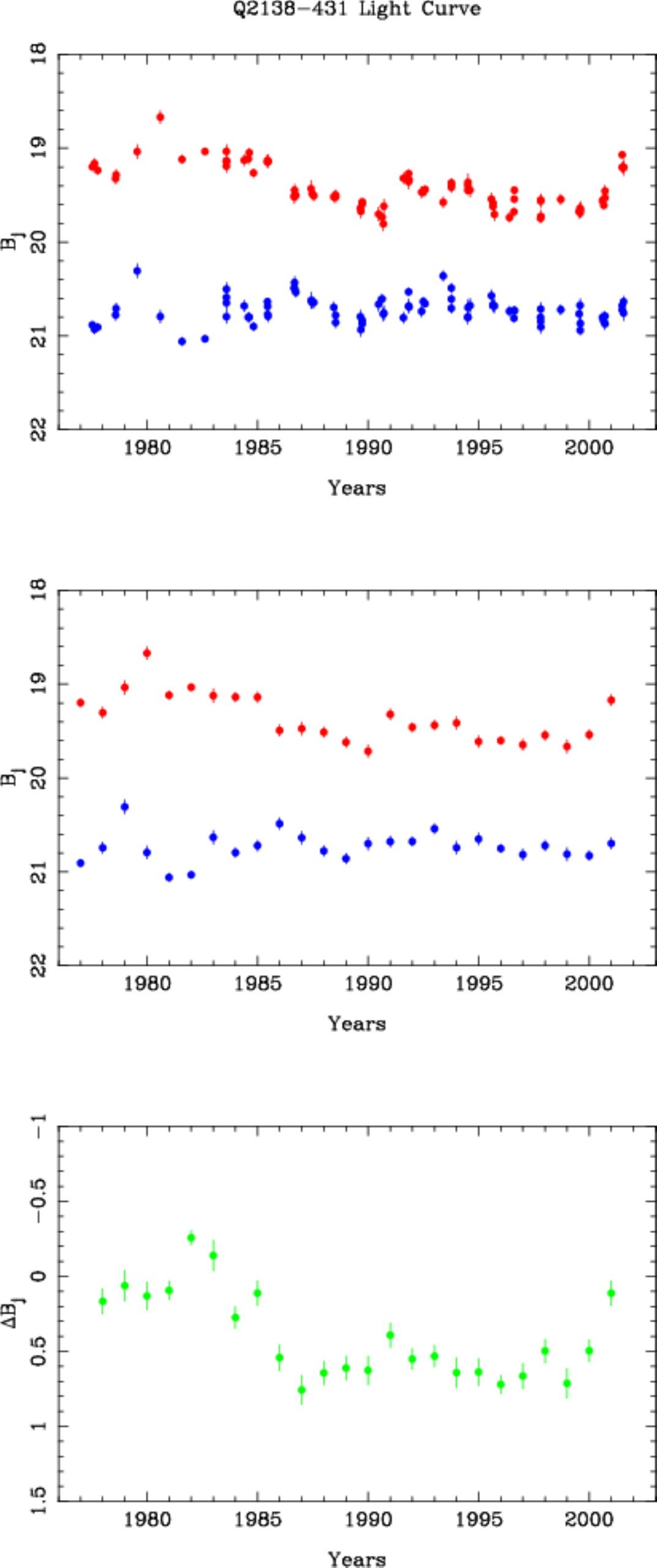}
\end{picture}
\caption{The top panel shows a 25 year light curve for image A (red)
 and image B (blue) of the double quasar Q2138-431, from SuperCOSMOS
 measures of a long series of UK 1.2m Schmidt plates in the $B_J$
 passband.  The middle panel shows the same data, binned in yearly
 intervals.  The bottom panel shows the difference light curve for the
 A and B images, with the observed lag of 1 year for image A removed.}
\label{fig5}
\end{figure}

The first stage in the calibration was to transform the transmision values
of the SuperCOSMOS scans to $I$, the intensity of light incident on the
photographic plate, using the relation:

\begin{equation}
I \propto \biggl({\frac{T_c}{T}}\biggr)^{1/\gamma}
\label{eqn3}
\end{equation}

\noindent where $T$ is the transmission through exposed emulsion, and
$T_c$ through clear plate.  $\gamma$ is the gradient of the linear part of
the response curve which in the first instance we set to 2.5.  However,
$\gamma$ is basically a scaling factor which becomes redundant with the
sequence star calibration.

The next step was to measure the magnitudes of the 18 sequence stars and
two quasar images on each of the 79 plate scans from the Supercosmos
archive which were available for incorporation into the light curves.
To do this, the transmision values were transformed to intensity using
equation~(\ref{eqn3}), and the resulting arrays converted to FITS files
for analysis with the GAIA photometry package.  Then for each plate the
sequence star and quasar images were measured with the aperture photometry
routine to give a quasi-linear relative flux, and thence converted to
instrumental magnitudes.  Using an aperture of radius 2 arcsec, the two
quasar images were easily resolved on all plates.  The final step was to
transform these instrumental magnitudes to the sequence star magnitudes
using a least squares third order polynomial fit.  This transformation was
then used to evaluate the true magnitudes of the two quasar images in the
$B_J$ passband.

The top panel of Fig.~\ref{fig5} shows the light curves for the two
quasar images from 1977 to 2001.  The error bars are derived from the
dispersion about the calibration curve, and are typically about 0.05
magnitudes, with little dependence on brightness.  The long term trend of
the light curves is better seen in the middle panel of Fig.~\ref{fig5}
where the observations for each year are averaged with a weighted mean.
Little is lost by this procedure, as in most years the plates were taken
within a period of about three months, and are not useful for measuring
changes within the year.  It will be seen that until 1985 the difference
between the two light curves averaged around 1.5 magnitudes.  After this
image A dimmed, and the difference decreased by about 0.3 magnitudes.
In the event that the double quasar is a gravitational lens system, we
would expect intrinsic variations in the quasar to show up in both light
curves, separated by a time interval corresponding to the difference in
light travel time to the two quasar images.  In addition, it is clear from
large scale monitoring programmes of gravitational lenses that in most
lensed systems the quasar images also vary independently of each other,
which is generally accepted to be the result of microlensing by stellar
mass bodies along the line of sight \citep{t13}.  To look for a time lag
between variations in the two images, and minimise the effects of
microlensing, we split the light curves into two sections defined by the
larger and smaller average magnitude differences, and measured the
correlation coefficient for different time lags.  The results are
displayed in Table~\ref{tab2}, and show for both time intervals a strong
correlation between the two light curves when image B leads image A by one
year.  Given the poor time resolution of the light curves, this result
should not be taken too seriously, but the existence of a time lag
provides confirmation that Q2138-431 is a gravitational lens, and is a
starting point for a more precise measurement.

\begin{table}
\caption{Correlation coefficients for time lags between images A and B of
 Q2138-431.}
\label{tab2}
\begin{center}
\vspace{5mm}
\begin{tabular}{lccccc}
\hline
 B leads A by (years) & 2 & 1 & 0 & -1 & -2 \\
\hline
 1977 -- 1985 & +0.04 & +0.70 & +0.04 & -0.69 & -0.59 \\
 1986 -- 1994 & -0.20 & +0.61 & +0.31 & +0.04 & +0.18 \\
\hline
\end{tabular}
\end{center}
\end{table}

The bottom panel of Fig.~\ref{fig5} shows the difference between the two
light curves in the middle panel, with the curve for image B shifted back
one year.  The zero-point for the difference curve can be calculated from
the magnitude difference of the two images in the infrared, where due to
the size of the emitting region the effects of microlensing are negligible
\citep{b11}.  From the $J$-band frames used for the detection of the
lensing galaxy we find for the magnitude difference
$J_{\rm A}-J_{\rm B}=-1.77$, which we use as the zero-point for the
difference curve in the bottom panel of Fig.~\ref{fig5}.  This now shows
the true differential effect of microlensing.

\section{Discussion}
\label{dis}

The idea behind this paper is to review the status of the double quasar
Q2138-431 by taking advantage of improved observational data since its
discovery in 1997.  At the time the case for classifying it as a
gravitational lens was very strong.  The spectra were almost identical,
as illustrated in Figures 1 and 2 of \cite{ha97}, and for two randomly
selected quasars this is very unlikely as illustrated in Figure 4 of
\cite{ha97}.  There is of course the possibility that two quasars evolving
in the same cluster environment could aquire the same observational
characteristics, but the actual mechanism by which this could happen
consistent with our knowledge of cluster evolution is far from clear.
The one sticking point in the way of classifying the system as a
gravitational lens was the failure to find a lensing galaxy.

Attempts to find a lensing galaxy by \cite{ha97} were mainly focussed on
deep CCD observations in the $R$-band.  They also reported observations
in the $K$-band, but their frame is of a similar depth to the $K_s$-band
frame in Fig.~\ref{fig2}, and barely detects image B of the quasar.  Their
$R$-band frame goes much deeper, and they derive a limit to the lensing
galaxy brightness of $R > 23.8$.  With this faint upper limit to the
brightness and an estimate of the lens mass they derive a very large lower
limit to the mass-to-light ratio of any lensing galaxy.  Their lens mass
is similar to the one we derive from mass modelling, but for reasons
which are not clear \cite{ha97} appear to assume zero redshift
luminosities, and do not take into account the very large $K$ and
evolutionary corrections associated with a distant galaxy.  When these are
allowed for, it seems unlikely that the galaxy we have detected in the
$J$-band would have been detected by \cite{ha97} in their $R$-band frame
for redshifts $z \gtrsim 0.6$.

Based on measurements of random fluctuations in the sky background, the
object revealed after PSF subtraction in Fig.~\ref{fig3} is a $5 \sigma$
detection.  Its appearance is similar to other faint galaxies in the
surrounding 10 arcmin field, where the presence of an optical counterpart
confirms their identity as galaxies.  A good example is the faint
galaxy to the south-east of the quasar system, used by \cite{ha97} to put
limits on the $R$-band magnitude of a lensing galaxy.  This galaxy is also
clearly visible on deep $r$- and $i$-band frames from the DES survey, and
is detected near the limit of the $J$-band frame in Fig.~\ref{fig3}.
There is in addition some indication of other faint galaxies in the
vicinity of image A of the quasar.  We have also searched the field for
artefacts which might be mistaken for faint galaxies, but these are
confined to single pixels, presumably cosmic ray tracks.

The value obtained for the mass-to-light ratio of the lens in
Section~\ref{gal} of $M_{\odot}/L_{\odot} = 9.0$ assumes a redshift of
$z = 0.6$ based on non-detection of the lens in the $i$-band.  As this
redshift is essentially a lower limit, we investigated the effect of
increasing it to $z = 1.0$ and derived $M_{\odot}/L_{\odot} = 8.2$ for
the lensing galaxy.  This small change in $M_{\odot}/L_{\odot}$ is due to
the fact that an increase in redshift results in larger values for both
luminosity and mass which tend to cancel out in the mass-to-light ratio.
Our value of $M_{\odot}/L_{\odot}$ lies quite close to the relation
between mass and mass-to-light found by \cite{c06}, and thus adds to a
consistent picture of the double quasar as a gravitational lens.

The mass modelling of Q2138-431 turned out to be straightforward.  This
was not unexpected, as the lensing galaxy lies close to the line of
centres of the quasar images, and roughly in the position one would expect
given their brightness ratio.  This contrasts strongly with for example the
asymmetric Q0957+561 \citep{f10}, and more closely resembles HE1104-1805
\citep{s12}, another wide separation binary.  This simplicity of modelling
is important if Q2138-431 is to be used for measuring the Hubble constant,
where the accuracy with which light travel time to the two images can be
modelled is the most important factor limiting the accuracy of the result.

The software package of \cite{k01} allows for the estimation of the time
delay between two images of the modelled system, assuming a value for the
Hubble constant.  We applied this procedure to our model of Q2138-431
using a canonical value $H_0 = 71$ km s$^{-1}$ Mpc$^{-1}$ which implied a
time lag with image B leading image A by 410 days.  This is consistent
with our crude estimate of 1 year from yearly observations, and hopefully
can provide a starting point for a well sampled photometric monitoring
programe designed to measure the value of $H_0$.  For this calculation we
assumed a lens redshift $z = 0.6$, equal to the lower limit derived in 
Section~\ref{gal}.  Increasing the lens redshift results in a larger
predicted time lag.  For example a lens redshift of $z=1.0$ implies a
time lag of 1130 days, which suggests that the true lens redshift is not
much greater than our lower limit of $z \gtrsim 0.6$.

In the discovery paper of the double quasar Q2138-431 \citep{ha97}, the
strongest evidence supporting the gravitational lens hypothesis was the
close similarity of the spectra of the two quasar images, and the very
small difference between their redshifts.  However, the authors concluded
that without the detection of a lensing galaxy the claim that the system
was a gravitational lens was insecure.  In this paper we have brought
together a number of new lines of argument to support the classification
of Q2138-431 as a gravitational lens.  The detection of the lensing galaxy
is perhaps the most conclusive, but the simplicity of the mass model, the
derivation of a plausible mass-to-light ratio, and the detection of a time
lag between the light curves of the two quasar images in agreement with
the mass model help to provide a consistent picture of a gravitational
lens system.  The measurement of the redshift of the lensing galaxy and
the accurate determination of the time lag between the two images should
then provide an excellent basis for a reliable estimate of the value of
the Hubble constant.

\section{Conclusions}

In this paper we have re-examined the question of whether the double
quasar Q2138-431 is a gravitational lens system.  The system was
discovered as part of a survey for quasars based on their variability,
and elongated images were included as possible detections of gravitational
lenses.  Early analysis of Q2138-431 \citep{ha97} showed it to comprise
two quasar images separated by 4.5 arcsec, with almost identical spectra
and redshifts.  This provided strong evidence for a gravitational lens,
but deep CCD photometry in the $R$-band failed to reveal a lensing galaxy.
The authors thus concluded that there was insufficient evidence to
definitively identify the system as a gravitational lens.

With the advent of more recent deep photometric surveys, especially in the
infrared, we searched again for a lensing galaxy.  In this case we
successfully detected a candidate lens on a deep $J$-band frame from the
VISTA Science Archive.  The apparent magnitude of the $5 \sigma$ detection
was measured to be $J = 20.68$, and non-detection in the optical bands
implied a redshift $z \gtrsim 0.6$.

The wide separation of the quasar images at 4.481 arcsec and the apparent
simplicity of the lens system make Q2138-431 an attractive candidate for
the measurement of the Hubble constant, which would require mass modelling
of the system.  Based on our measurements of the positions of the quasar
images and lensing galaxy we were able to obtain a satisfactory fit with a
${\rm SIE}+\gamma$ model, implying an estimated mass for the lens of
$m_{lens} = 1.31 \times 10^{12} M_{\odot}$.  Combining this with the
infrared photometry in the $J$-band gives a mass-to-light ratio
$M_{\odot}/L_{\odot} = 9.0$ for the lensing galaxy.

The estimation of the Hubble constant in a gravitational lens system also
requires a measurement of the time delay or difference in light travel
time between the quasar images.  We were able to make a preliminary
assessment of this from an archival photographic monitoring programme
covering 21 years.  Although the effective time resolution of the survey
was only about 1 year, we were able to show by cross-correlating the light
curves of the two quasar images that image B leads image A by about a
year.  It was possible to estimate from our mass model that the expected
time delay was about 410 days, consistent with the rough value from the
light curve analysis.

The overall conclusion of this paper is that the double quasar Q2138-431
is confirmed as a gravitational lens system.  The close similarity of the
spectra, the detection of the lensing galaxy, the plausible mass-to-light
ratio, and the measurement of a time delay between the two images in
agreement with predictions from the mass model provide a consistent
picture of a gravitational lens.  Q2138-431 appears to be a system well
suited for the measurement of the Hubble constant.  The wide separation of
the quasar images should make measuring the time delay from photometry of
the light curves straightforward, and the simplicity of the system should
enable accurate modelling to estimate the value of the Hubble constant.

\section*{Acknowledgements}

I am most grateful to Nigel Hambly for retrieving the SuperCOSMOS scans
used for constructing the light curves.

\section*{Data Availability}

The data upon which this paper is based are all publicly available and
are referenced in Section~\ref{obs}, with some additional comments in the
text.

\bsp	
\label{lastpage}
\end{document}